\documentclass[12pt]{article}

\usepackage{amsmath}
\usepackage{amssymb}
\usepackage[dvips]{graphicx}
\usepackage[unicode=true]{hyperref}

\newcommand{\N}{N\raise.7ex\hbox{\underline{$\circ $}}$\;$}
\begin{document}

\title{On Some Discrete Subgroups of the Lorentz Group}
\author{A.N. Tarakanov
\thanks{E-mail: tarak-ph@mail.ru }\\
{\small Institute of Informational Technologies,}\\ {\small
Belarusian State University of Informatics and Radioelectronics}\\
{\small Kozlov str. 28, 220037, Minsk, Belarus}}
\date{}
\maketitle

\begin{abstract}
Some discrete subgroups of the Lorentz group are found using
Fedorov's parame\-trization by means of complex vector-parameter.
It is shown that the discrete subgroup of the Lorentz group, which
have not fixed points, are contained in boosts along a spatial
direction for time-like and space-like vectors and are discrete
subgroups of the group $SO(1,1)$, whereas discrete subgroups of
isotropic vector are subgroups of $SO(1,1) \times E(1,1)$.
\end{abstract}

 \vskip 5mm

 \vskip 25mm

From the physical point of view discrete subgroups of the Lorentz
group arise when one attempts to construct a theory of quantized
space-time with some discrete symmetry going over to Lorentz
symmetry at continual limit (see, e.g.,~\cite{Pot}). Under such
discrete transformation the space-time, represented as some
1+3-dimensional lattice, should go over into itself. Thus, the
problem is to find these discrete transformations, which,
obviously, should belong to discrete subgroup of the Lorentz
group. Despite numerous approaches to construction of
1+3-dimensional lattices, this problem remains unresolved till now
though there is some advancement in this direction (see,
e.g.,~\cite{Mak}-~\cite{Bal}). Works~\cite{Dir},~\cite{Sch} should
also be noted where some discrete subgroups of the Lorentz group
are constructed starting from homomorphism between $SO(1,3)$ and
$SL(2,C)$. The invariance principle under such subgroups, which
act independently on the particle states with various momenta,
allows defining all elements of the S-matrix~\cite{Bel}.

The purpose of this work is to set an example of construction of
discrete subgroups of the Lorentz group on the basis of chosen
parametrization. For construction of discrete subgroups we will
use Fedorov's parametrization of the Lorentz group by means of
complex vector-parameter $\mathbf{q} = \mathbf{a} + i
\mathbf{b}$~\cite{Fed}. Let us give the underlying information.

As it is known, the Lorentz group is a group of motions of the
Minkowski space $\mathbf{E}^{R}_{1,3}$. It will be a discrete
point group of symmetry if two conditions are satisfied: a) there
exists at least one point called singular one, which is invariant
under all transformations of group; b) the orbit of any
nonsingular point is discrete (~\cite{Bal}, p. 94). If
$\mathbf{L}(q) \in SO(1,3)$, $\mathrm{x} \in
\mathbf{E}^{R}_{1,3}$, then the first condition implies
$\mathbf{L}(q) \mathrm{x} = \mathrm{x}$, which thus selects the
little Lorentz group from the whole group. The second condition
specifies lattice in the Minkowski space.

The matrix of the Lorentz transformation is given by
$$
 \mathbf{L}(\mathrm{q}) = \frac{(1+ \mathrm{q})(1 +
 \mathrm{q}^{*})}{|1 + \mathbf{q}^2|} \equiv
 \mathbf{L}(\mathbf{q}) =
$$
$$
 = \frac{1}{|1 + \mathbf{q}^2|} \left (\begin{array}{cc}
                                  1 + |\mathbf{q}|^2 & i(\mathbf{q} - \mathbf{q}^{*} + [\mathbf{q} \mathbf{q}^{*}]) \\
                                  i(\mathbf{q} - \mathbf{q}^{*} - [\mathbf{q} \mathbf{q}^{*}]) & 1 - |\mathbf{q}|^2 + (\mathbf{q} + \mathbf{q}^{*})^{\times} + \mathbf{q} \circ \mathbf{q}^{*} + \mathbf{q}^{*} \circ \mathbf{q}\\
                                \end{array} \right)
  \;, \eqno{(1)}
$$
where $4\times4$-matrix $\mathrm{q}$ has the form
$$
 \mathrm{q} = \left (\begin{array}{cc}
                                  0 & i\mathbf{q} \\
                                  i\mathbf{q} & \mathbf{q}^{\times} \\
                                \end{array} \right)
  \;, \; \;
  \mathrm{q}^{*} = \left (\begin{array}{cc}
                                  0 & -i\mathbf{q}^{*} \\
                                  -i\mathbf{q}^{*} & \mathbf{q}^{*\times} \\
                                \end{array} \right)
  \;, \eqno{(2)}
$$
$3\times3$-matrix $\mathbf{q}^{\times}$ has components
$(\mathbf{q}^{\times})_{ij} = \varepsilon_{ijk} q_k$, and the sign
$\circ$ implies dyadic product: $(\mathbf{q} \circ
\mathbf{q}^{*})_{ij} = q_{i} q^{*}_{j}$. Matrix (1) satisfies to
condition of pseudo-orthogonality
$$
\mathbf{L}(\mathrm{q}) \mathrm{\eta}
\widetilde{\mathbf{L}}(\mathrm{q}) = \mathrm{\eta} \;, \; \;
\mathrm{\eta} = \left (
\begin{array}{cc}
                                                   1 & 0 \\
                                                   0 & -\mathbf{1} \\
                                                 \end{array}
                                                 \right )
 \;. \eqno{(3)}
$$

The composition law of vector-parameters looks like
$$
\mathbf{q}'' = <\mathbf{q}, \mathbf{q}'>\equiv \frac{\mathbf{q} +
\mathbf{q}' + [\mathbf{q} \mathbf{q}']}{1 - \mathbf{q}\mathbf{q}'}
 \;, \eqno{(4)}
$$
and $\mathbf{L}(\mathbf{q}'') = \mathbf{L}(\mathbf{q})
\mathbf{L}(\mathbf{q}')$. Let's write down an action of
$\mathbf{L}(\mathbf{q})$ on a vector $\mathrm{x} = \left
(\begin{array}{c}
                                                    x^0 \\
                                                     \mathbf{x} \\
                                                   \end{array} \right )$:
                                                   $\mathrm{x}' = \mathbf{L}(\mathbf{q})
                                                   \mathrm{x}$, or
$$
\left\{ \begin{array}{c}
         'x_0 = \dfrac{1}{|1 + \mathbf{q}^2|} \{(1 + |\mathbf{q}|^2) x^{0} +i(\mathbf{q} - \mathbf{q}^{*} + [\mathbf{q} \mathbf{q}^{*}]) \mathbf{x}\} \;, \\
         '\mathbf{x} = \dfrac{1}{|1 + \mathbf{q}^2|} \{i(\mathbf{q} - \mathbf{q}^{*} - [\mathbf{q} \mathbf{q}^{*}])x^{0} + (1 - |\mathbf{q}|^2) \mathbf{x} + [(\mathbf{q} + \mathbf{q}^{*}), \mathbf{x}] + \mathbf{q} (\mathbf{q}^{*} \mathbf{x}) + \mathbf{q}^{*} (\mathbf{q} \mathbf{x})\} \;. \\
       \end{array}\right. \; \eqno{(5)}
$$

Matrix $\mathbf{L}(\mathbf{q})$ can be represented also as

$$
\mathbf{L}(\mathbf{q}) = \frac{1 + \boldsymbol{\alpha}}{1 -
\boldsymbol{\alpha}} = \frac{1 +\frac{1}{2}
(\mathbf{q}^{2}+\mathbf{q}^{*2}) + 2 (\boldsymbol{\beta} +
\boldsymbol{\beta}^{2})}{|1 + \mathbf{q}^2|} \;, \eqno{(6)}
$$
where $\boldsymbol{\alpha}$ is anti-Hermitian matrix:
$\boldsymbol{\alpha}^{\dag} = -\boldsymbol{\alpha}$. Its relation
with vector-parameter looks as follows:
$$
\boldsymbol{\alpha} = \xi \boldsymbol{\beta} + \zeta
\boldsymbol{\beta}^{3} \;, \eqno{(7)}
$$
$$
 \boldsymbol{\beta} = \frac{1}{2} (\mathbf{q} + \mathbf{q}^{*}) = \left (\begin{array}{cc}
                                                                    0 & -\mathbf{b} \\
                                                                    -\mathbf{b} & \mathbf{a}^{\times} \\
                                                                  \end{array}
                                                                  \right )
 \;, \; \boldsymbol{\beta}^{\dag} = -\boldsymbol{\beta} \;, \eqno{(8)}
$$
$$
\xi = 1 - \Delta_{\beta} \frac{\sqrt{(1 - \Delta_{\beta})^{2} - 4
|\boldsymbol{\beta}|} - (1 - \Delta_{\beta})}{2
|\boldsymbol{\beta}|} \;, \; \zeta = \frac{\sqrt{(1 -
\Delta_{\beta})^{2} - 4 |\boldsymbol{\beta}|}}{2
|\boldsymbol{\beta}|} \;, \eqno{(9)}
$$
where
$$
\Delta_{\beta} = \frac{1}{2} \mathrm{Sp}{\boldsymbol{\beta}^2} =
-\frac{1}{2} (\mathbf{q}^{2} + \mathbf{q}^{*2}) \;, \;
|\boldsymbol{\beta}| = \mathrm{det}{\boldsymbol{\beta}} =
\frac{1}{16} (\mathbf{q}^{2} - \mathbf{q}^{*2})^{2} \;.
\eqno{(10)}
$$

Further we will need to know the structure of small Lorentz group
leaving a vector x fixed: $\mathbf{L}(\mathbf{q}) \mathrm{x} =
\mathrm{x}$. Here we have three cases that we will consider below.

1. $\mathrm{x}$ is \textit{a time-like vector}, $(x^0)^{2} -
\mathbf{x}^{2} > 0$. In this case by means of some transformation
$\mathbf{L}(\mathbf{c})$ one may obtain a vector
$\stackrel{\circ}{\mathrm{x}} = (x^{0}, \mathbf{0})$

$$
 \stackrel{\circ}{\mathrm{x}} = \mathbf{L}(\mathbf{c}) \mathrm{x}
 \;,\eqno{(11)}
$$

Using properties of the little Lorentz group we obtain from
Eq.(11)
$$
\stackrel{\circ}{\mathrm{x}} = \mathbf{L}(\mathbf{c}) \mathrm{x} =
\mathbf{L}(\mathbf{c}) \mathbf{L}(\mathbf{q}) \mathrm{x} =
\mathbf{L}(\mathbf{c}) \mathbf{L}(\mathbf{q})
\mathbf{L}(-\mathbf{c}) \stackrel{\circ}{\mathrm{x}} =
\mathbf{L}(<\mathbf{c}, \mathbf{q}, -\mathbf{c}>)
\stackrel{\circ}{\mathrm{x}} = \mathbf{L}(O(\mathbf{c})
\mathbf{q}) \stackrel{\circ}{\mathrm{x}} \;, \eqno{(12)}
$$
where
$$
O(\mathbf{c}) = \mathbf{1} + 2 \frac{\mathbf{c}^{\times} +
(\mathbf{c}^{\times})^2}{1 + \mathbf{c}^2} \eqno{(13)}
$$
is a matrix from the complex rotation group. Thus, the vector
$\stackrel{\circ}{\mathrm{x}}$ will not change under Lorentz
transformation $\mathbf{L}(\mathbf{q}')$ with the vector-parameter
$$
\mathbf{q}' = O(\mathbf{c}) \mathbf{q} \;. \eqno{(14)}
$$
The possible structure of the vector $\mathbf{q}'$ is specified
from equations (5), where $\mathbf{q}$ and $\mathrm{x}'$  should
be replaced by $\mathbf{q}'$ and $\stackrel{\circ}{\mathrm{x}}$,
respectively. Then these equations give
$$
\stackrel{\circ}{x^0} = \frac{1 + |\mathbf{q}'|^2}{|1 +
\mathbf{q}'^2} x^{0} \;, \eqno{(15)}
$$
$$
\mathbf{q}' - \mathbf{q}'^{*} - [\mathbf{q}' \mathbf{q}'^{*}] =
\mathbf{0} \;. \eqno{(16)}
$$
Hence it follows $\mathbf{q}' = \mathbf{q}'^{*}$ and
$\mathbf{L}(\mathbf{q}')$ takes the form
$$
\mathbf{L}(\mathbf{q}') = \left ( \begin{array}{cc}
                            1 & 0 \\
                            0 & O(\mathbf{q}') \\
                          \end{array} \right )\;, \eqno{(17)}
$$
i.e. it is a 3-dimensional rotation.

Using for $\mathbf{L}(\mathbf{q})$ a representation (6), we obtain
for the little group of the vector $\mathrm{x}$
$$
\boldsymbol{\alpha} \mathrm{x} = (\xi \boldsymbol{\beta} + \zeta
\boldsymbol{\beta}^{3}) \mathrm{x} = \left (\begin{array}{cc}
                                     0 & -\mathbf{B} \\
                                             -\mathbf{B} & \mathbf{A}^{\times} \\
                                           \end{array} \right ) \left (\begin{array}{c}
                                                                  x^0 \\
                                                                  \mathbf{x} \\
                                                                \end{array}
                                                                \right
                                                                )
                                                                =
                                                                \left
                                                                (\begin{array}{c}
                                                                   -\mathbf{B} \mathbf{x} \\
                                                                   -\mathbf{B} x^{0} + [\mathbf{A} \mathbf{x}] \\
                                                                 \end{array}
                                                                 \right
                                                                 )
                                                                 =
                                                                 0 \;, \eqno{(18)}
$$
where
$$
\mathbf{A} = (\xi - \zeta a^{2} + \zeta b^{2}) \mathbf{a} - \zeta
(\mathbf{a}\mathbf{b}) \mathbf{b} \;, \eqno{(19)}
$$
$$
\mathbf{B} = (\xi - \zeta a^{2} + \zeta b^{2}) \mathbf{b} + \zeta
(\mathbf{a}\mathbf{b}) \mathbf{a} \;. \eqno{(20)}
$$

It follows from Eq.(18) that vectors $\mathbf{A}$ and $\mathbf{B}$
are orthogonal: $(\mathbf{A}\mathbf{B}) = 0$, but it is fulfilled
only at $|\boldsymbol{\beta}| = 0$. Then $\boldsymbol{\alpha} =
\boldsymbol{\beta}$ and $\mathbf{A} = \mathbf{a}$, $\mathbf{B} =
\mathbf{b}$. As it is shown in~\cite{Bea}, $|\boldsymbol{\beta}| =
0$ is necessary and sufficient condition for matrix
$\mathbf{L}(\mathbf{q})$ to have eigenvalue 1 that takes place for
the little Lorentz group. Thus, vectors $\mathbf{a}$ and
$\mathbf{b}$ are orthogonal, and vector $\mathbf{b}$ is orthogonal
to $\mathbf{x}$
$$
(\mathbf{a}\mathbf{b}) = 0 \;, \; (\mathbf{b}\mathbf{x}) =0 \;.
\eqno{(21)}
$$
The first condition means the vector $\mathbf{q}$ to be canonical.
When the vector $\stackrel{\circ}{\mathrm{x}}$ is time-like,
Eq.(21) is fulfilled automatically due to $\mathbf{b}' = 0$. In
general case of time-like vector $\mathrm{x}$ a transformation
from the little Lorentz group is given by
$$
\mathbf{L}(\mathbf{q}) = \mathbf{L}(O(-\mathbf{c}) \mathbf{a}')
\;, \eqno{(22)}
$$
where $O(-\mathbf{c}) = O^{-1}(\mathbf{c})$ is specified in
Eq.(13). The vector $\mathbf{q} = \mathbf{a} + i\mathbf{b}$
satisfies to conditions (21), implying $\mathbf{q}$ to be
represented as
$$
\mathbf{q} = \mathbf{a} + i \varepsilon [\mathbf{e}_{1}
\mathbf{a}] = \mathbf{a} - \frac{i \mathbf{x}}{x^0}
[\mathbf{e}_{1} \mathbf{a}] \;, \eqno{(23)}
$$
where $\mathbf{e}_{1}$ is a unit vector in the
$\mathbf{x}$-direction: $\mathbf{e}_{1} =
\frac{\mathbf{x}}{|\mathbf{x}|}$, $\varepsilon =
-\frac{|\mathbf{x}|}{x^0}$. The structure of vectors $\mathbf{q}$
and $\mathbf{q}'$ determines the structure of the vector
$\mathbf{c}$:
$$
\mathbf{c} = \mu (\mathbf{q} + \mathbf{q}') + \frac{2 [\mathbf{q}
\mathbf{q}']}{(\mathbf{q} + \mathbf{q}')^2} \;. \eqno{(24)}
$$
From $\mathbf{q}^{2} = \mathbf{q}'^2$ it follows $(\mathbf{e}_{1}
\mathbf{a})^{2} = \mathbf{a}'^2$. Substituting $\mathbf{q}' =
\mathbf{a}'$ in Eqs.(23) and (24), we obtain
$$
\mathbf{c} = \mu \left(\mathbf{a} + \mathbf{a}' -
\frac{i[\mathbf{x} \mathbf{a}]}{x^0} \right) + \frac{[\mathbf{a}
\mathbf{a}'] + \frac{i}{x^0} (\mathbf{x} [\mathbf{a} \mathbf{a}']
- \mathbf{a}' (\mathbf{x} \mathbf{a}))}{\mathbf{a}' (\mathbf{a} +
\mathbf{a}') - \frac{i (\mathbf{x} [\mathbf{a}
\mathbf{a}'])}{x^0}} \;, \eqno{(25)}
$$
where $\mu$ is arbitrary complex number.

2. $\mathrm{x}$ is \textit{a space-like vector}, $(x^0)^{2} -
\mathbf{x}^{2} < 0$. In this case some transformation
$\mathbf{L}(\mathbf{d})$ can give a vector
$\stackrel{\circ}{\mathrm{x}} = (0, \stackrel{\circ}{\mathbf{x}})$
$$
 \stackrel{\circ}{\mathrm{x}} = \mathbf{L}(\mathbf{d}) \mathrm{x}
 \;.\eqno{(26)}
$$
For such $\stackrel{\circ}{\mathrm{x}}$ we also have relations
(12)-(14), where $\mathbf{c}$ is replaced by $\mathbf{d}$. For
arbitrary space-like vector $\mathrm{x}$ conditions (18), where
$\mathbf{A} = \mathbf{a}$ and $\mathbf{B} = \mathbf{b}$, and (21)
are fulfilled. For the vector $\stackrel{\circ}{\mathrm{x}}$
Eqs.(18), (21) transform into
$$
[\mathbf{a}' \stackrel{\circ}{\mathbf{x}}] = \mathbf{0} \;, \;
(\mathbf{a}'\mathbf{b}') = 0 \;, \; (\mathbf{b}'
\stackrel{\circ}{\mathbf{x}}) = 0 \;. \eqno{(27)}
$$
i.e. vector $\mathbf{a}'$ is parallel to
$\stackrel{\circ}{\mathbf{x}}$, and vector $\mathbf{b}'$ is
orthogonal to $\stackrel{\circ}{\mathbf{x}}$ and $\mathbf{a}'$.
Replacing $\mathbf{q}$ by $\mathbf{q}'$ in Eq.(5), and
$\mathrm{x}'$ by space-like $\stackrel{\circ}{\mathrm{x}}$, we
obtain
$$
(\mathbf{q}' - \mathbf{q}'^{*} + [\mathbf{q}' \mathbf{q}'^{*}])
\stackrel{\circ}{\mathbf{x}} = 0 \;. \eqno{(28)}
$$
$$
\stackrel{\circ}{\mathbf{x}} = \frac{1}{|1 + \mathbf{q}'^2|} \left
\{(1 - |\mathbf{q}'|^{2}) \stackrel{\circ}{\mathbf{x}} +
[(\mathbf{q}' + \mathbf{q}'^{*}), \stackrel{\circ}{\mathbf{x}}] +
\mathbf{q}' (\mathbf{q}'^{*} \stackrel{\circ}{\mathbf{x}}) +
\mathbf{q}'^{*} (\mathbf{q}' \stackrel{\circ}{\mathbf{x}}) \right
\} \;, \eqno{(29)}
$$
or
$$
\left\{ \begin{array}{c}
         (\mathbf{b}' \stackrel{\circ}{\mathbf{x}}) - (\stackrel{\circ}{\mathbf{x}} [\mathbf{a}' \mathbf{b}']) = 0 \;, \\
         \mathbf{a}'^{2} \stackrel{\circ}{\mathbf{x}} - [\mathbf{a}' \stackrel{\circ}{\mathbf{x}}] - \mathbf{a}' (\mathbf{a}' \stackrel{\circ}{\mathbf{x}}) - \mathbf{b}' (\mathbf{b}' \stackrel{\circ}{\mathbf{x}}) = \mathbf{0} \;. \\
       \end{array}\right. \; \eqno{(30)}
$$
The first condition is fulfilled identically, whereas from the
second one it follows
$$
\mathbf{a}' = a'
\frac{\stackrel{\circ}{\mathbf{x}}}{|\stackrel{\circ}{\mathbf{x}}|}
= a' \stackrel{\circ}{\mathbf{e}}_{1} \;. \eqno{(31)}
$$

Let us choose as unit vectors of the basis the vectors
$$
\stackrel{\circ}{\mathbf{e}}_{1} \;, \;
\stackrel{\circ}{\mathbf{e}} \; = \frac{1}{\sqrt{2}}
(\stackrel{\circ}{\mathbf{e}}_{2} + i
\stackrel{\circ}{\mathbf{e}}_{3}) \;, \;
\stackrel{\circ}{\mathbf{e}}^{*} = \frac{1}{\sqrt{2}}
(\stackrel{\circ}{\mathbf{e}}_{2} - i
\stackrel{\circ}{\mathbf{e}}_{3}) \;, \eqno{(32)}
$$
satisfying to algebra
$$
[\stackrel{\circ}{\mathbf{e}}_{1}, \stackrel{\circ}{\mathbf{e}}] =
-i \stackrel{\circ}{\mathbf{e}} \;, \;
[\stackrel{\circ}{\mathbf{e}}_{1},
\stackrel{\circ}{\mathbf{e}}^{*}] = i
\stackrel{\circ}{\mathbf{e}}^{*} \;, \;
[\stackrel{\circ}{\mathbf{e}}, \stackrel{\circ}{\mathbf{e}}^{*}] =
-i \stackrel{\circ}{\mathbf{e}}_{1} \;; \eqno{(33)}
$$
$$
(\stackrel{\circ}{\mathbf{e}}_{1})^{2} = 1 \;, \;
(\stackrel{\circ}{\mathbf{e}})^{2} =
(\stackrel{\circ}{\mathbf{e}}^{*})^{2} = 0 \;, \;
\stackrel{\circ}{\mathbf{e}} \stackrel{\circ}{\mathbf{e}}^{*} =1
\;; \stackrel{\circ}{\mathbf{e}}_{1} \stackrel{\circ}{\mathbf{e}}
\; = \; \stackrel{\circ}{\mathbf{e}}_{1}
\stackrel{\circ}{\mathbf{e}}^{*} = 0 \;. \eqno{(34)}
$$
Then the vector-parameter $\mathbf{q}'$ can be represented as
$$
\mathbf{q}' = a' \stackrel{\circ}{\mathbf{e}}_{1} + i (\gamma'
\stackrel{\circ}{\mathbf{e}} + \gamma'^{*}
\stackrel{\circ}{\mathbf{e}}^{*}) \;, a' = a'^{*} \;, \eqno{(35)}
$$
and $\mathbf{L}(\mathbf{q}')$ looks like
$$
\mathbf{L}(\mathbf{q}') = \frac{1}{1 + a'^{2} - 2 \gamma'
\gamma'^{*}} \left( \begin{array}{cc}
  1 + a'^{2} + 2 \gamma' \gamma'^{*}  & - 2 \gamma'^{*} (1-ia') \stackrel{\circ}{\mathbf{e}}^{*} - \\
                                      &  - 2 \gamma' (1-ia') \stackrel{\circ}{\mathbf{e}} \\
  - 2 \gamma' (1-ia') \stackrel{\circ}{\mathbf{e}} - & 1 - a'^{2} - 2 bb^{*} + \\
  - 2 \gamma'^{*} (1+ia') \stackrel{\circ}{\mathbf{e}}^{*} & + 2a' \stackrel{\circ}{\mathbf{e}}_{1}^{\times} + 2 a'^{2} \stackrel{\circ}{\mathbf{e}}_{1} \circ \stackrel{\circ}{\mathbf{e}}_{1} + \\
                                                           & + 2 (\gamma' \stackrel{\circ}{\mathbf{e}} + \gamma'^{*} \stackrel{\circ}{\mathbf{e}}^{*}) \circ (\gamma' \stackrel{\circ}{\mathbf{e}} + \gamma'^{*} \stackrel{\circ}{\mathbf{e}}^{*})
\end{array} \right). \eqno{(36)}
$$

In the general case of space-like vector x a transformation from
the little Lorentz group is given by
$$
\mathbf{L}(\mathbf{q}) = \mathbf{L}(O(-\mathbf{d}) \mathbf{q}')
\;, \eqno{(37)}
$$
where vector $\mathbf{q}$, specified in the same way as in
time-like case in Eq.(23), can be written down also in the form
$$
\mathbf{q} = a \mathbf{e}_{1} + \left (1 -
\frac{|\mathbf{x}|}{x^0} \right ) b \mathbf{e} + \left (1 +
\frac{|\mathbf{x}|}{x^0} \right ) b^{*} \mathbf{e}^{*} \;,
\eqno{(38)}
$$
and $\mathbf{e}_{1}$, $\mathbf{e}$, $\mathbf{e}^{*}$ satisfy the
same relations as Eqs.(33), (34), and $\mathbf{a} = a
\mathbf{e}_{1} + b \mathbf{e} + b^{*} \mathbf{e}^{*}$. The
connection between $\mathbf{q}$, $\mathbf{q}'$, and $\mathbf{d}$
looks like
$$
a \mathbf{e}_{1} + \left (1 - \frac{|\mathbf{x}|}{x^0} \right ) b
\mathbf{e} + \left (1 + \frac{|\mathbf{x}|}{x^0} \right ) b^{*}
\mathbf{e}^{*} =
$$
$$
= \left (1 + 2\frac{-\mathbf{d}^{\times} +
\mathbf{d} \circ \mathbf{d} - \mathbf{d}^{2}}{1 + \mathbf{d}^{2}}
\right ) (a' \stackrel{\circ}{\mathbf{e}}_{1} + i \gamma
\stackrel{\circ}{\mathbf{e}} + i \gamma^{*}
\stackrel{\circ}{\mathbf{e}}^{*}) \;, \eqno{(39)}
$$
where it should be keep in mind that $\mathbf{e} = \;
\stackrel{\circ}{\mathbf{e}}$, $\mathbf{e}^{*} = \;
\stackrel{\circ}{\mathbf{e}}^{*}$, and
$\stackrel{\circ}{\mathbf{e}}_{1} \; =
\frac{\stackrel{\circ}{\mathbf{x}}}{|\stackrel{\circ}{\mathbf{x}}|}$
is determined from Eq.(26), or
$$
\stackrel{\circ}{\mathbf{x}} = \frac{(\mathbf{d} - \mathbf{d}^{*}
- [\mathbf{d} \mathbf{d}^{*}]) ((\mathbf{d} - \mathbf{d}^{*} +
[\mathbf{d} \mathbf{d}^{*}])\mathbf{x})}{(1 + |\mathbf{d}|^{2}) |1
+ \mathbf{d}^{2}|} +
$$
$$
+ \frac{(1 - |\mathbf{d}|^{2}) \mathbf{x} + [(\mathbf{d} +
\mathbf{d}^{*}), \mathbf{x}] + \mathbf{d} (\mathbf{d}^{*}
\mathbf{x}) + \mathbf{d}^{*} (\mathbf{d} \mathbf{x})}{|1 +
\mathbf{d}^{2}|} \;. \eqno{(40)}
$$
It is seen from here, that sought connection between $\mathbf{q}$,
$\mathbf{q}'$ and $\mathbf{d}$ is not so simple as in Eq.(24).

3. x is \textit{isotropic vector}, $(x^0)^{2} - \mathbf{x}^{2} =
0$. In this case an explicit form of the vector-parameter
$\mathbf{q}$ follows from Eq.(23):
$$
\mathbf{q} = \mathbf{a} - i [\mathbf{e}_{1} \mathbf{a}] \;.
\eqno{(41)}
$$
Hence a transformation from the little Lorentz group takes the
form
$$
\mathbf{L}(\mathbf{q}') = \frac{1}{1 + (\mathbf{e}_{1}
\mathbf{a})^2} \left(
\begin{array}{cc}
  1 + 2a^{2} - (\mathbf{e}_{1} \mathbf{a})^2 & -2 (a^{2} \mathbf{e}_{1} - \mathbf{a} (\mathbf{e}_{1} \mathbf{a}) - [\mathbf{e}_{1} \mathbf{a}]) \\
  2 [a^{2} \mathbf{e}_{1} - \mathbf{a} (\mathbf{e}_{1} \mathbf{a}) + [\mathbf{e}_{1} \mathbf{a}]] & 1 - (\mathbf{e}_{1} \mathbf{a})^{2} + 2 \mathbf{a}^{\times} - 2a^{2} \mathbf{e}_{1} \circ \mathbf{e}_{1} + \\
                                                                                                  & + 2 (\mathbf{e}_{1}\mathbf{a})(\mathbf{e}_{1} \circ \mathbf{a} + \mathbf{a} \circ \mathbf{e}_{1})
\end{array}
\right ). \eqno{(42)}
$$
If vector $\mathbf{a}$ is orthogonal to $\mathbf{x}$-direction,
i.e. $(\mathbf{e}_{1} \mathbf{a}) = 0$, we have
$$
\mathbf{L}(\mathbf{q}') = \left(
\begin{array}{cc}
  1 + 2a^{2} & -2 [a^{2} \mathbf{e}_{1} - [\mathbf{e}_{1} \mathbf{a}]] \\
  2 [a^{2} \mathbf{e}_{1} + [\mathbf{e}_{1} \mathbf{a}]] & 1 + 2 \mathbf{a}^{\times} - 2a^{2} \mathbf{e}_{1} \circ \mathbf{e}_{1}
\end{array}
\right ). \eqno{(43)}
$$
When $\mathbf{a}$ is parallel to $\mathbf{x}$, then $\mathbf{q} =
\mathbf{a}$ and $\mathbf{L}(\mathbf{q}')$ looks like Eq.(17).

Let us consider now discrete Lorentz transformations. Let
components of the vector-parameter $\mathbf{q}$ be rational
complex numbers, i.e. numbers with real and imaginary parts
looking like $m/n$, where $m$ and $n$ are integers. Then a
composition law (4) of two rational vector-parameters generates
rational vector-parameters as well. Identity and inverse elements
correspond to $\mathbf{q} = \mathbf{0}$ and $\mathbf{q}' = -
\mathbf{q}$, respectively. It is obvious they are also rational.
Matrix $\mathbf{L}(\mathbf{q})$ specified in Eq.(1) is not
rational, for $|1 + \mathbf{q}^2| = \sqrt{(1 + \mathbf{q}^2)(1 +
\mathbf{q}^{*2})}$ is irrational in general. Nevertheless its
components accept discrete set of values, specified by
discreteness of rational values of the vector-parameter
$\mathbf{q}$. Thus, rational $\mathbf{q}$'s give discrete
subgroups of the Lorentz group. Setting some initial coordinates
$\mathrm{x} = (x^{0}, \mathbf{x})$ which are not necessarily
possessing property of rationality, by means of Lorentz
transformations we will obtain new coordinates $\mathrm{x}' =
\mathbf{L}(\mathbf{q}) \mathbf{x}$, and, enumerating all possible
rational values of $\mathbf{q}$, we will obtain some discrete set
of points. Obviously, we cannot obtain such set if
$\mathbf{L}(\mathbf{q})$ belongs to the little Lorentz group
leaving points $\mathrm{x}$ immovable. Thus, \textit{in order that
the discrete subgroup of the Lorentz group did not contain
elements which leave vectors immovable, it should not contain
discrete subgroups of the little Lorentz group}.

The little Lorentz group is $SO(3)$ for time-like vectors,
$SO(1,2)$ for space-time vectors, and a group isomorphic to group
$E(2)$ of flat motions, for vector (41) can be represented in the
form
$$
\mathbf{q} = a_{1} \mathbf{e}_{1} + 2 a^{*} \mathbf{e}^{*} =
<a_{1} \mathbf{e}_{1}, \frac{2 a^{*}}{1 + ia_1} \mathbf{e}^{*}>
\;, \eqno{(44)}
$$
with
$$
\mathbf{a} = a_{1} \mathbf{e}_{1} + a \; \mathbf{e} + a^{*}
\mathbf{e}^{*} \;, \; a_{1} = a_{1}^{*} \;. \eqno{(45)}
$$
The vector-parameter $a_{1} \mathbf{e}_{1}$ corresponds to
rotation on an angle $\varphi = 2 \arctan{a_1}$ round the
$\mathbf{x}$-direction, and vector-parameter $\frac{2 a^{*}}{1 +
ia_1} \mathbf{e}^{*}$ corresponds to translation in the plane
which is orthogonal to $\mathbf{x}$. Hence, subgroups of the
Lorentz group not having immovable points are contained in boosts,
generating groups $SO(1,1)$, along the $\mathbf{x}$-direction for
time-like and space-time vectors. In the case of isotropic vectors
such subgroups are contained in the group generated by the
vector-parameter
$$
\mathbf{q} = i b \mathbf{e}_{1} + c \; \mathbf{e} = <i b
\mathbf{e}_{1}, \frac{c}{1 + b} \mathbf{e}> \;. \eqno{(46)}
$$
The vector-parameter $i b \mathbf{e}_{1}$ corresponds to boosts,
generating group $SO(1,1)$, in the $\mathbf{x}$-direction, and
vector $\frac{c}{1 + b} \mathbf{e}$ gives rise simultaneously to
(various) dilatations of temporal coordinate and vector
$\mathbf{x}$ and translations in the plane which is orthogonal to
$\mathbf{x}$. Denoting this group as $E(1,1)$ one may assert that
discrete subgroups of isotropic vector are subgroups of the group
$SO(1,1) \times E(1,1)$, where symbol $\times$ means semidirect
product.

Let us consider boosts $SO(1,1)$ in the $\mathbf{x}$-direction,
which are inherent to all kinds of vectors and specified by the
vector-parameter
$$
\mathbf{q} = i b \mathbf{e}_{1} \;. \eqno{(47)}
$$
A composition law (4) reduces to composition of parameter $b$:
$$
b'' = \frac{b + b'}{1 + bb'} \;. \eqno{(48)}
$$

Here one can see at least two types of discrete subgroups.

1) $b$ is rational number, $b = m/n$;

2) $b = \tanh{(\mu r)}$, where $r = m/n$ is an integer or rational
number, which composition

\hskip 5mm law is trivial: $r'' = r + r'$; $0 < \mu < \infty$ is a
fixed real number, determining continuum

\hskip 5mm of discrete subgroups of such kind. Here the subgroup
is extracted corresponding

\hskip 5mm to integer $r$.

For the group $E(1,1)$ specified by the vector-parameter
$$
\mathbf{q} = d \mathbf{e} = \frac{c}{1+b} \mathbf{e} \;,
\eqno{(49)}
$$
a composition law (4) reduces to
$$
d'' = \frac{d + d'}{1 - dd'} \;, \; \; \mathrm{or} \; \;
\frac{c''}{1 + b''} = \frac{c(1 +b') + c'(1 + b)}{(1+b)(1+b') -
cc'} \;. \eqno{(50)}
$$

Here we also obtain two two types of discrete subgroups.

1) $d$ is rational number, $d = m/n$;

2) $d = \tan{(\mu r)}$, where $r = m/n$ is an integer or rational
number, which composition

\hskip 5mm law is trivial: $r'' = r + r'$; $0 < \mu < \infty$ is a
fixed real number, determining continuum

\hskip 5mm of discrete subgroups of such kind. Here the subgroup
is also extracted correspond-\

\hskip 5mm ing to integer $r$.

For the group $SO(1,1) \times E(1,1)$, taking into account the law
(48) in Eq.(50), we obtain a composition law for $c$:
$$
c'' = \frac{c(1 + b') + c'(1 + b)}{(1+bb') \left [1 -
\frac{cc'}{(1 + b)(1 + b')} \right ]} \;. \eqno{(51)}
$$
From here it follows

1) if $b = m/n$, then $c$ is also rational: $c = p/q$, with $d =
\frac{k}{l} = \frac{pn}{q(m+n)}$ and
$$
\frac{p''}{q''} = \frac{[pq'(n' + m') + p'q(n + m)](n + m)(n' +
m')}{[qq'(n + m)(n' + m') - nn'pp'] (nn' + mm')} \;; \eqno{(52)}
$$

2) if $b = \tanh{(\mu m/n)}$, $d = \tanh{(\nu k/l)}$, then $c =
\tanh{(\nu k/l)} [1 + \tanh{(\mu m/n)}]$.

In conclusion it should be noted that the Lorentz group may be
parametrized by various ways, which determine range of parameters.
Hence, discrete subgroups may be obtained in various
parametrizations. A question about whether arbitrary
parametrization admits existence of discrete subgroups of the
Lorentz group, by now remains opened.


\begin{thebibliography}{99}

\bibitem{Pot} Potter F. Unification of Interactions in Discrete
Spacetime. // Progr. in Phys., 2006, \textbf{1}, 3-9.

\bibitem{Mak} Makarov V.S. Geometric methods of construction of
the discrete groups of motion of the Lobatchevsky's space. // In:
Problemy geometrii. Itogi nauki I techniki. 1983. Vol. 15. -- pp.
3-59 (in Russian).

\bibitem{Apa1} Apanasov B.N. Discrete groups of transformations
and structures of manifolds. -- Novosibirsk: 1983; 2nd ed. -- M.:
Nauka, 1991. -- 426 pp. (in Russian).

\bibitem{Apa2} Apanasov B.N. Geometriya diskretnyh grupp i
mnogoobrazij. -- M.: Nauka, 1991. -- 214 pp. (In Russian).

\bibitem{Bea} Beardon A.F. The geometry of discrete groups
(Graduate Texts in Mathematics, \textbf{91}). 2nd ed. -- New York:
Springer, 1983. -- xii+356 pp.

\bibitem{Bal} Baltag I.A. Metody postroyeniya discretnych grupp
preobrazovaniy simmetrii prostranstva Minkowskogo. -- Chisinau:
Stiinta, 1987. -- 196 pp. (in Russian).

\bibitem{Dir} Dirac P.A.M. Discrete subgroups of the Poincar\'{e}
group. // In: Problemi teoretichestoi fiziki, Ed. V.I.Ritus. --
Moscow: Publ. House Nauka, 1972. -- pp. 45-51.

\bibitem{Sch} Schwarz F. On Discrete Subgroups of the Lorentz
Group. // Lett. Nuovo Cim., 1976, \textbf{15}, no. 1, 7-14.

\bibitem{Bel} Belavin A.A. Discrete groups and the integrability
of quantum systems. // Functional Analysis and Its Applications,
1980, \textbf{14}, no. 4, 260-267.

\bibitem{Fed} Fedorov F.I. Lorentz Group. -- Moscow: Nauka, 1979.
M. -- 385 pp. (in Russian).

\end{thebibliography}
\end{document}